\documentclass[sigconf]{acmart}
\usepackage{CJK}  
\usepackage{algorithm}  
\usepackage{algorithmicx}  
\usepackage{algpseudocode}  
\usepackage{amsmath}  
\usepackage{amssymb}  
\usepackage{geometry} 
\usepackage{url}

\usepackage{tikz}
\usepackage{pgfplots}

\usepackage{graphicx}
\usepackage{epstopdf}

\usepackage{multirow}

\graphicspath{{./Figures/}}

\setcopyright{none} 

\begin{document}
\title{Catering to Your Concerns: Automatic Generation of Personalised Security-Centric Descriptions for Android Apps} 

\author{Tingmin Wu}
\affiliation{
	\institution{Swinburne University of Technology}}
\email{tingminwu@swin.edu.au}

\author{Lihong Tang}
\affiliation{
	\institution{Swinburne University of Technology}}
\email{lihongtang@swin.edu.au}

\author{Rongjunchen Zhang}
\affiliation{
	\institution{Swinburne University of Technology}}
\email{rongjunchenzhang@swin.edu.au}

\author{Sheng Wen}
\affiliation{
	\institution{Swinburne University of Technology}}
\email{swen@swin.edu.au}

\author{Cecile Paris}
\affiliation{
	\institution{Data61 CSIRO, Australia}}
\email{Cecile.Paris@data61.csiro.au}

\author{Surya Nepal}
\affiliation{
	\institution{Data61 CSIRO, Australia}}
\email{Surya.Nepal@data61.csiro.au}

\author{Marthie Grobler}
\affiliation{
	\institution{Data61 CSIRO, Australia}}
\email{Marthie.Grobler@data61.csiro.au}

\author{Yang Xiang}
\affiliation{
	\institution{Swinburne University of Technology}}
\email{yxiang@swin.edu.au}

\begin{abstract}
Android users are increasingly concerned with the privacy of their data and security of their devices. To improve the security awareness of users, 
recent automatic techniques produce security-centric descriptions by performing program analysis. However, the generated text does not always address users' concerns as they are generally too technical to be understood by ordinary users. Moreover, different users have varied linguistic preferences, which do not match the text. Motivated by this challenge, we develop an innovative scheme to help users avoid malware and privacy-breaching apps by generating security descriptions that explain the privacy and security related aspects of an Android app in clear and understandable terms. We implement a prototype system, PERSCRIPTION, to generate personalised security-centric descriptions that automatically learn users' security concerns and linguistic preferences to produce user-oriented descriptions. We evaluate our scheme through experiments and user studies. The results clearly demonstrate the improvement on readability and users' security awareness of PERSCRIPTION's descriptions compared to existing description generators.  
\end{abstract}



\maketitle

\section{Introduction}
With the phenomenal growth of the Android platform, there has been a tremendous increase in security threats. The openness of the environment attracts criminals to spread malicious apps such as Trojan \cite{zhou2012hey}, Worms \cite{mulliner2011poster}, Botnets \cite{pieterse2012Android} and Spyware \cite{seo2014detecting}. Malicious activities can lead to privilege escalation, privacy leaks, personal information theft and monetary loss \cite{faruki2015Android}.

To assist users in deciding whether they should install an app, Android provides two kinds of information: requested permissions and textual descriptions generated by app developers. It is vital that end users understand the potential security risks and make the right decision (\textit{e.g.}, avoiding installing malicious apps) based on the information provided. However, it has been shown that this information cannot effectively protect users from installing malware or privacy-breaching apps \cite{zhang2015towards}. In general, researchers have already demonstrated that the requested permissions and textual descriptions do not significantly improve security awareness of end users \cite{zhang2015towards}. 
Request permissions are used to declare sensitive resources (\textit{e.g.}, camera, location, contacts) used by apps. This is in part because the majority of end users do not understand the currently used technical permission terms, limiting the readability of the permission lists.
Hence, the provided information is limited for the purpose of decision making. A user study shows that a substantial number of users are unaware of the presence of the permission list \cite{felt2012Android}. Thus, there is a high possibility that users still tend to accept malicious app installation without having a full understanding of security implications from the permission list. In addition, app developers usually sidestep security risks in their textual descriptions. More importantly, variations exist between the produced texts and the real functionality of apps. Criminals can easily craft deceptive textual app descriptions to mislead inexperienced users. A significant deviation is also found between the requested permissions and the provided descriptions \cite{qu2014autocog}.

To address the aforementioned problem, textual descriptions generated from app code analysis are considered a feasible alternative solution. Suspicious codes are detected by performing program analysis, and then the outcomes are transformed into human readable language through NLG (Natural Language Generation); example systems in this category include DescribeMe \cite{zhang2015towards}, AutoPPG \cite{arp2014drebin}, and Drebin \cite {yu2017toward}. Although the descriptions generated by these systems are security-centric, they are still too technical and tedious to attract users' attention to improve their security awareness. The overall impact of such systems are no better than permission lists since the descriptions are likely to be neglected by end users \cite{felt2012Android}.  

In this paper, we propose an innovative method that automatically learns users' security concerns and linguistic preferences to generate personalised security-centric descriptions that can improve their security awareness. We learn users' concerns through the ratio of the permissions that they modify deliberately. To learn linguistic preferences, we first classify individuals into distinct personality types based on their behaviours, and then produce corresponding linguistic preferences. The NLG technique is integrated to organise and generate personalised text that can be easily understood by users so that they can pay the necessary attention.

We implement a prototype system, PERSCRIPTION, to demonstrate the novelty of our approach. We leverage machine learning algorithms in user personality prediction, and utilise Statistical Natural Language Generation (SNLG) to generate stylistic variations in textual descriptions for different personalities. We evaluate our system using real-world Android apps from different app markets. Experiment results and user studies demonstrate the improvement of our generated security descriptions compared to recent approaches \cite{zhang2015towards,arp2014drebin} in terms of readability and increased user awareness. It is also demonstrated that our system is efficient to generate semantically-correct results by consuming low energy in a short time, which is an extremely important requirement for mobile devices. To the best of our knowledge, we are the first to develop a psychologically-informed security-centric description generator for the Android platform. We leverage the Big Five personality traits \cite{john1999big} to classify Android users according to the adoption of different types of mobile apps. The Big Five model is exploitd in our NLG method as human perceptions of personality variations are delineated by the model \cite{pennebaker1999linguistic}. The words associated with each personality trait are demonstrated to represent a specific linguistic style \cite{mairesse2011controlling}. Some studies reveal the correlation between personality and linguistic cues \cite{alam2014fusion,majumder2017deep,pennebaker2001linguistic}.

In summary, we make the following contributions:
\begin{description}
\item[$\bullet$]We propose an innovative method that automatically learns users' concerns and linguistic preferences, and generates personalised security-centric descriptions to improve their security awareness. 
\item[$\bullet$]We implement our prototype system, PERSCRIPTION, by incorporating state-of-the-art techniques from different disciplines, such as Big Five personality traits classification, linguistic style generation via NLG, and generating psychologically-informed descriptions.
\item[$\bullet$]Experiments results and user studies show significant improvement of our generated descriptions in terms of readability and user awareness compared to other state-of-the-art techniques.
\end{description}

The rest of this paper is structured as follows. Section \ref{s_overview} presents the overview of PERSCRIPTION. Section \ref{s_con}, \ref{s_lin} and \ref{s_NLG} detail three parts of our system with a proof-of-concept respectively. Experiment results are described in Section \ref{s_evaluation}. We discuss the limitations and related work in Section \ref{s_dis} and Section \ref{s_relatedwork} respectively, and we conclude the paper in Section \ref{s_conclusion}.

\section{Overview}
\label{s_overview}
\subsection{Problem Statement}
\begin{figure}[t]
\centering
\includegraphics[width=1\linewidth]{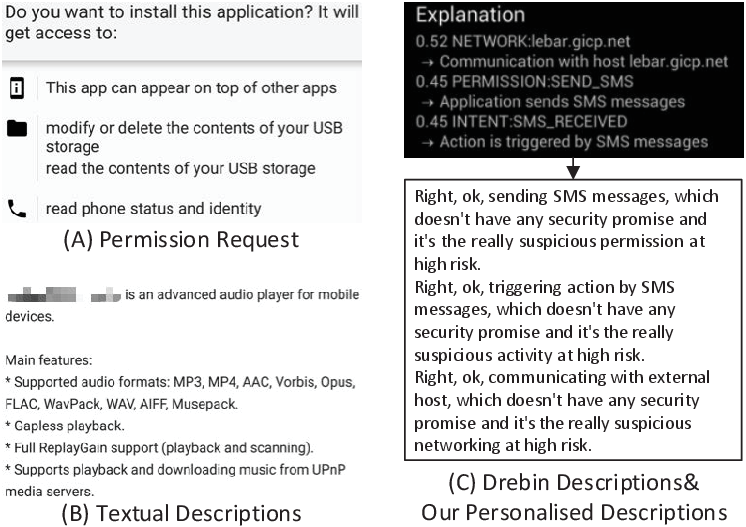}
\caption{Existing descriptions v.s. our personalised descriptions before app installation.}
\label{f_ExampleDescription}
\end{figure}

Figure \ref{f_ExampleDescription} depicts the current methods for security description generation. Figure \ref{f_ExampleDescription}A provides an example of requested permission list including reading phone information. The simple enumeration does not always get users' attention and protect users from malicious app installation. The longer the enumeration list, the harder it is to attract users' attention. Similarly, the textual descriptions are generally generated from app functionality or users' instructions instead of users' security concerns. An example is provided in Figure \ref{f_ExampleDescription}B. Significant deviation is also revealed between permission lists and textual descriptions \cite{qu2014autocog}.

\begin{figure}[t]
\centering
\includegraphics[width=1\linewidth]{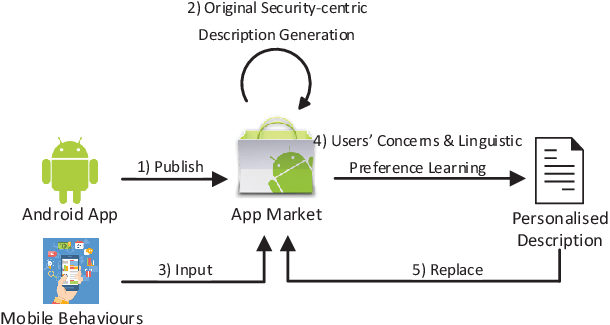}
\caption{Deployment of PERSCRIPTION.}
\label{f_Deployment}
\end{figure}

Recently, some efforts have been made in explainable detection on malicious Android apps \cite{arp2014drebin,zhang2015towards,yu2017toward}. These methods extract suspicious codes by performing program analysis and craft the textual descriptions using NLG. Consider Drebin \cite{arp2014drebin} as an example (the top part of Figure \ref{f_ExampleDescription}C), it detects the most representative features of the malware and inserts them into several simple sentence templates as security descriptions. There is no user-centric method to generate personalised security descriptions for different users.

To address this issue, we propose a new technique, PERSCRIPTION, with automatic linguistic stylistic description generation that address users' concerns psychologically. A sample of our personalised descriptions is shown at the bottom of Figure \ref{f_ExampleDescription}C. It demonstrates a learned linguistic style for one personality dimension with trained parameters from statistical models. As shown in Figure \ref{f_ExampleDescription}C, the content of our scripts is the same as Drebin, while the advantage of our descriptions is noticeable in terms of naturalness, readability, and usefulness. We believe our method is more effective in assisting users in enhancing awareness of security threats in an unknown app. From our experiment results, 25\% more users were aware of the threats through our descriptions compared to Drebin \cite{arp2014drebin}.

The expected deployment model of our system is demonstrated in Figure \ref{f_Deployment}. After the developer publishes an app in an app market, the original security-centric descriptions will be generated automatically by performing code analysis. When a user attempts to install the app, the user's mobile information is obtained automatically to personalise app security descriptions with specific content and linguistic style suitable for the specific user. The personalised descriptions are shown on mobile screens instead of the original ones.

The aim of PERSCRIPTION is to generate personalised descriptions that cater to users' concerns. This goal is achieved as follows. 1) \textbf{Security-Centric Description.} Our provided descriptions are security-centric as they are derived from the malicious patterns (malware indicative features) detected by existing method \cite{arp2014drebin}. 2) \textbf{Personalised Description.} The generated descriptions target users' concerns precisely, with the related part automatically lifted to the top in their preferred linguistic style. 
3) \textbf{Natural and Readable Description.} The output scripts are natural and readable with stylistic linguistic expression to match users with different personalities.

\begin{figure*}
\centering
\includegraphics[width=1\linewidth]{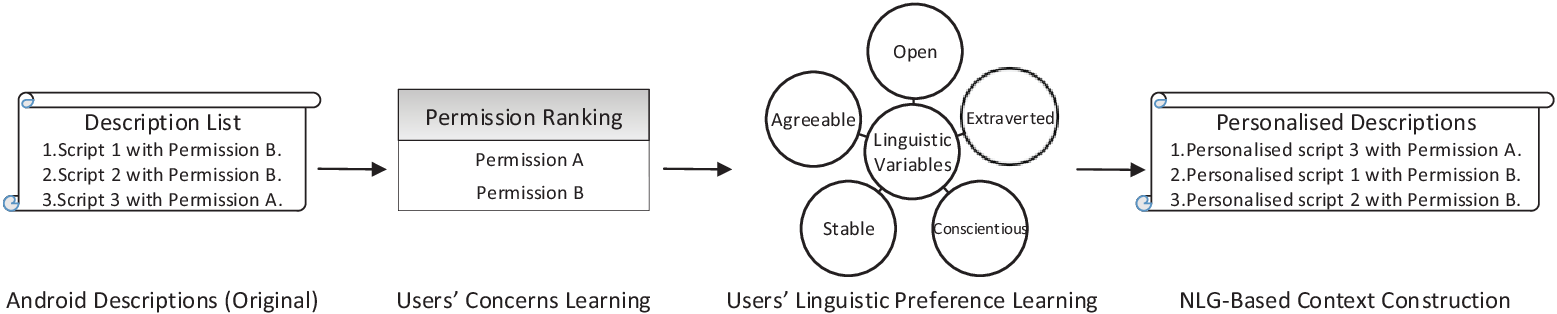}
\caption{Overview of PERSCRIPTION.}
\label{f_overview}
\end{figure*}

\subsection{Architecture Overview}
Our system workflow is illustrated in Figure \ref{f_overview}. The personalised description generator includes the following three steps:

1) \textbf{Users' Concerns Learning.} We study users' concerns according to their mobile permission change. Intuitively, frequent default setting modifications indicate users' concerns on different permissions. For instance, if a user refuses to provide Location permission for the listed apps, Location is then considered as the permission user concerns. In this case, the proportion of denied apps represents the extent of the user's concerns about Location. Based on the extents of different permissions, the descriptions are then reorganised to be consistent with the sorting result.

2) \textbf{Users' Linguistic Preference Learning.} To generate personalised descriptions for different users, we classify users into different types to design unique text for each of them. The users are characterised into different personalities based on the Big Five Personality model \cite{john1999big}. Mobile behaviours (app adoption) are used as indicators to predict their personalities using machine learning algorithms. Linguistic preference (preferred words) for each personalty is then learned for the next step.
 
3) \textbf{NLG-Based Context Construction.} Finally, we construct the text according to the users' concerns and linguistic preference. In the process, the syntactic structures are generated for malicious information detected from an app. They are later realised as sentences to be organised and aggregated into one security-centric description. To generate scripts for different personalities of users, Statistical Natural Language Generation (SNLG) is employed, and a series of stylistic factors and parameters (\textit{e.g.}, aggregation operations and pragmatic markers) are determined by learning algorithms to demonstrate continuous variation in linguistic preference. For example, introverts are more likely to adopt complex structures \cite{mairesse2011controlling}.

\subsection{Baseline}
As our work mainly focuses on description comprehension and reconstruction, we apply Drebin \cite{arp2014drebin} as the malware detection baseline. Drebin extracted thousands of features from app code and manifest files such as API calls and permission request. The features were utilised to train a malware detection model using machine learning algorithms. Representative features for malicious apps were also learned from the training model. The features were meaningful to provide effective descriptions to help users to consider security threats.

Drebin then fed the obtained features into eight simple pre-designed sentence templates without considering human readability and awareness. Figure \ref{f_ExampleDescription}C (top part) presents Drebin descriptions for a malware sample. Although the malicious behaviours were accurately detected including communication with external server and sending SMS messages, the risk of security threats was hardly perceived by humans with the fixed-template text generation. This becomes the motivation of our research.

\begin{figure}[t]
\centering
\includegraphics[width=1\linewidth]{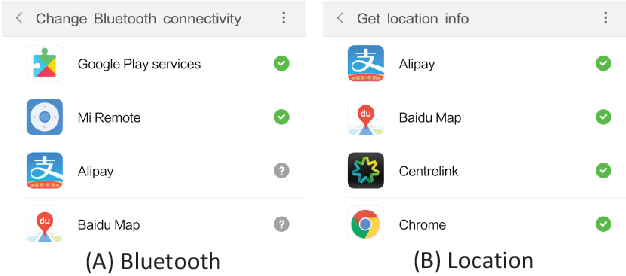}
\caption{Example permission layouts for (A) Bluetooth and (B) Location.}
\label{f_permissionLayout}
\end{figure}

\section{Users' Concerns Learning}
\label{s_con}
In this section, we learn users' concerns from their behaviours on app requested permissions. In total, we apply eight representative permissions in our experiment: Location, Contacts, Calendars, Reminders, Photos, Bluetooth, Microphone and Camera. They are displayed as several individual layouts in mobile devices. Two examples (Bluetooth and Location) are shown in Figure \ref{f_permissionLayout}. Each layout contains an app list which requires the permission at installation time. The green button (with check mark) along with the app name represents the access status of `Allow', whereas the grey button (with question mark) indicates `Deny'. In the provided examples, all apps are allowed to access the Location. Users can modify the status to enable (`Allow') or disable (`Deny') any app to access the related permission after installation.
\subsection{Permission Preference Learning}
We learn users' permission preferences which indicate their privacy concerns. According to the Android Document \cite{Android-doc}, app developers need to state the required permissions in <uses-permission> elements of manifest files that users must grant to run the app correctly.  Hence, the access statuses for app requested permissions are set as `Allowed' in default. Some prudent users modify them to prevent their personal information from leaking. We access the app list with statuses for each permission from the mobile device and label the status as modified if it is `Deny'. If a user changes the statuses substantially in some specific permissions, the permissions are considered as the users' concerns. More concretely, we define a variable `Attention Level' (a percentage ranged from 0 to 1) to represent the extent of their concerns on each permission. Specifically, the variable is calculated as the proportion of the modified statuses in the app list for each permission. Afterwards, the ranking of 8 permissions are generated ordered by their 'Attention Level'.

\begin{figure}[t]
\centering
\includegraphics[width=1\linewidth]{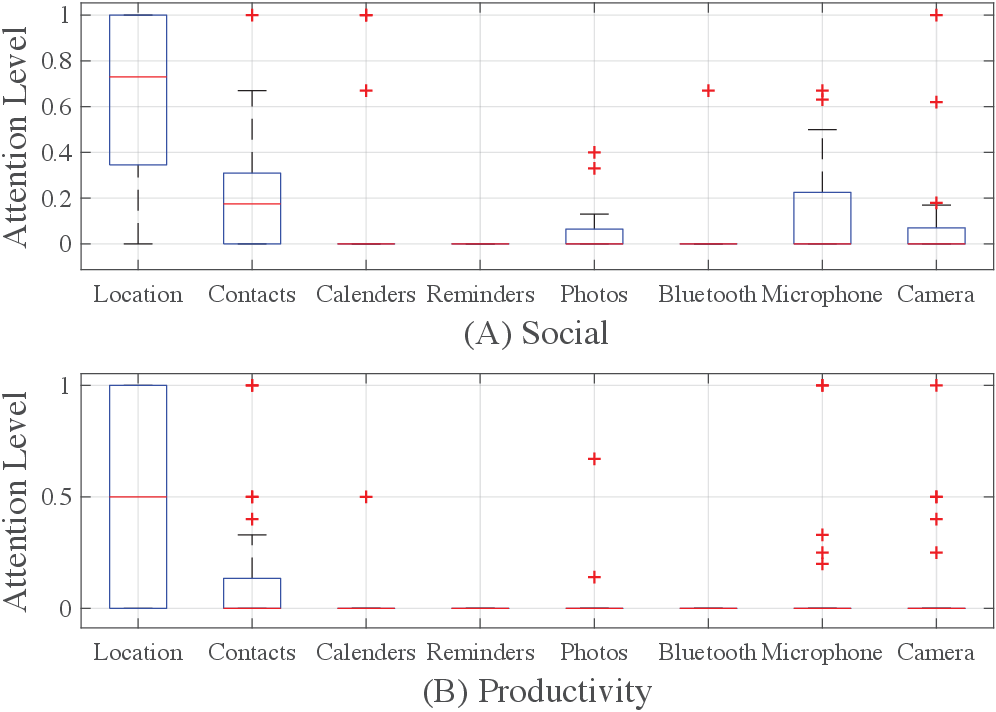}
\caption{Users' Attention Levels in different app categories: (A) Social and (B) Productivity. The box represents the Attention Level distribution in 36 participants from our user study in Section \ref{ss_poc1}. Red lines presents the medians. Data points beyond the whiskers are displayed using `+'.}
\label{f_userConcernSpec}
\end{figure}

\begin{figure}[t]
\centering
\includegraphics[width=1\linewidth]{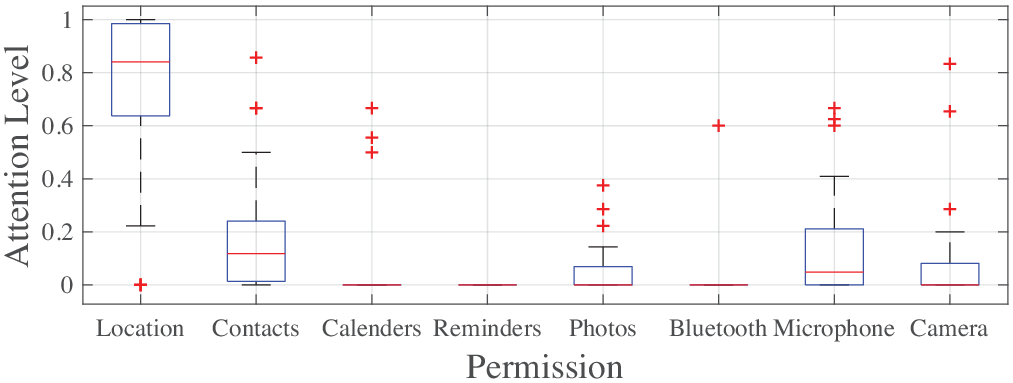}
\caption{Attention Level distribution on different permissions.}
\label{f_userCatering}
\end{figure}

Moreover, users' concerns vary in different types of apps, such that users may concern more about the permission Location on the app `Google Maps' than `News'. Therefore, we further classify apps into eight categories (Audio, Game, Image, Maps, News, Productivity, Social and Video) according to Android Document \cite{Android-doc}. We learn a user's concerns in each category separately to provide the specific permission ranking based on the category of the app to be installed. Figure \ref{f_userConcernSpec} shows a example of users' Attention Levels distributed on eight permissions for two app categories: Social and Productivity. It is indicated that permission rankings can be different on different categories.



Algorithm \ref{a_permission} demonstrates our permission ranking generation at app installation time. The inputs of the algorithm are the targeted category of the app to be installed, the set of installed apps and the applied eight permissions. The first step is to traverse all installed apps to select those in the targeted category and add them into a specific group. If the user has not installed any app in the mobile before, the default ranking is provided which is the statistical result from other users. If there are no installed apps in the targeted category, the remaining apps are applied in filling the group. The proportion of the modified settings in the specific app list is computed to represent the Attention Level for each permission separately. Finally, the permissions are sorted accordingly in descending order.



\label{ss_permissionSorting}
        \begin{algorithm}[t]
            \caption{Permission Ranking Generation at App Installation}  
            \begin{algorithmic}[1] 
            \Require $cate$: the category of the app to be installed, $Set_{app}$: a set of installed apps, $Set_{per}$: a set of permissions
            \Ensure  $R_{per}$: permission rankings for the app to be installed
            
            \State $R_{per} = \emptyset$;
            \State $Set_{specapp} = \emptyset$;
            \If {Exist($Set_{app}$)}
                \State $Set_specapp \gets \{app \in cate\} $;
            \Else
                \State \Return $defaultRanking$;
            \EndIf
            \If {!Exist($Set_{specapp}$)}
                \State $Set_{specapp} \gets Set_{app}$;
            \EndIf
            \State $R_{per} \gets \{GetAttentionLevel(Set_{specapp} \in Set_{per}\})$;
            \State Sort($R_{per}$);
            \State \Return $R_{per}$;
                
            \end{algorithmic}  
            \label{a_permission}
        \end{algorithm} 
        
\subsection{Description Reconstruction}
The descriptions from Drebin \cite{arp2014drebin} are reconstructed based on the permission rankings. As the permissions indicate users' privacy concerns, we reorder the sentences conforming to the rankings. For example, if the permission Location is ranked first, the related descriptions will be lifted to the top. The purpose of the design is to attract users' attention to the security threats when they attempt to download a new app. The order of the generated sentences is important, because most users only spend a moment to read the top part of the text. If they are attracted at a glance, it is more likely for them to read the rest and further realise the serious security risks from the app.

\subsection{Proof-of-Concept: Part I}
\label{ss_poc1}

\subsubsection{Data Collection}
We conduct a user study to collect real-world user permission settings. Before we start, we obtained approval of our submitted full ethics application from the Human Ethics Advisory Groups. We selected mobile users as our target group, and randomly targeted pedestrians at different places such as universities or shopping centres. The collection process is challenging. Some people are concerned about privacy leak by providing such private information, even with our promise of confidentiality protection. In addition, each participant is required to give the full app lists along with the associated eight permissions. In total, we obtained valid information from 36 volunteers. To guarantee the correctness, we asked them to capture snapshots containing required permission settings and send the pictures to us. We further manually classify the apps from the pictures into defined categories and record the settings for them (\textit{i.e.}, Allow or Deny).
    
\subsubsection{User Permission Analysis}
It is depicted in Figure \ref{f_userCatering} that a user's concerns on different permissions vary greatly. We calculate the ratio of the apps with setting change in the whole list to represent their concern levels for each permission. Figure \ref{f_userCatering} shows the distribution of concern levels for all participants. It can be found that the level of concerns on Location is the highest among the users, and the level ranges around 62\% to nearly 100\%. On the contrary, the concern levels for all the remaining permissions are lower than 25\%. It is indicated that most individuals fail to consider the risks in three permissions: Calendars, Reminders and Bluetooth.

\begin{table}[t]
\caption{Big Five dimensions with six specific facets \cite{mccrae1992introduction}.}
\tiny
\label{t_bigFiveDimension}
\centering
\begin{tabular}{l l l l l}
\hline
E & A & C & N & O \\ \hline
Extraversion & Agreeableness & Conscientiousness & Neuroticism & Openness \\ \hline
Gregariousness & Trust & Competence & Anxiety & Ideas \\ 
Assertiveness & Dutifulness & Order & Angry hostility & Fantasy \\ 
Activity & Altruism & Dutifulness & Depression & Aesthetics \\ 
Excitement-seeking & Compliance & Achievement striving & Self-consciousness & Actions \\ 
Positive emotions & Modesty & Self-discipline & Impulsiveness & Feelings \\ 
Warmth & Trust & Deliberation & Vulnerability & Values \\ \hline
\end{tabular}
\end{table}

\section{Users' Linguistic Preferences Learning}
\label{s_lin}
To learn linguistic variation for different users, we first build a classifier to predict user traits. We then study the distinct linguistic preferences in each trait including text organisation and pragmatic markers.

\subsection{Big Five Personality Trait}
\label{ss_personality}
The Big Five model \cite{john1999big} is widely accepted in psychology to precisely address users' personality. We therefore use it as a bridge between users' mobile behaviours and linguistic preferences. As shown in Table \ref{t_bigFiveDimension}, the Big Five personality traits include Extraversion (E), Agreeableness (A), Conscientiousness (C), Neuroticism (N) and Openness (O). The six facets that define each dimension \cite{mccrae1992introduction} are also listed in the table. For instance, people in extraversion are found to be more talkative, outgoing, active and sociable compared to other dimensions. 

A classic measurement, Big Five Inventory (BFI), is used to identify users' personality traits in Big Five dimensions \cite{john1999big}. BFI contains 44 questions, and each of them demonstrates a representative characteristic. To complete the questionnaire, users are required to rate each characteristic on a sliding scale, where 1 represents \textit{Disagree Strongly} and 5 represents \textit{Agree Strongly}. The overall scores in the five dimensions are calculated according to the scoring instructions \cite{john1999big}, and each dimension is calculated using the sum of scores with its related questions. It is indicated from prior studies \cite{codish2014personality,xu2016understanding} that there is no standard way to determine whether the dimension score is high or low. To solve this issue, we borrow the result of a reliable user study including more than 70,000 individuals \cite{BFI_Manual}. Each dimension is labelled as high or low according to the comparison between calculated score and the corresponding mean value from the result. There exists a variation on the mean values in male and female. Hence, demographics information is also required to determine the labels for ground truth data. However, the studies \cite{maccallum2002practice,xu2016understanding} demonstrate the shortage of dichotomizing quantitative variables, and the fact that only a small percentage of people perform as high or low in each personality trait. Therefore, we employ three groups (High, Medium, Low) to label participants in five traits separately.

\begin{figure}[t]
\centering
\includegraphics[width=1\linewidth]{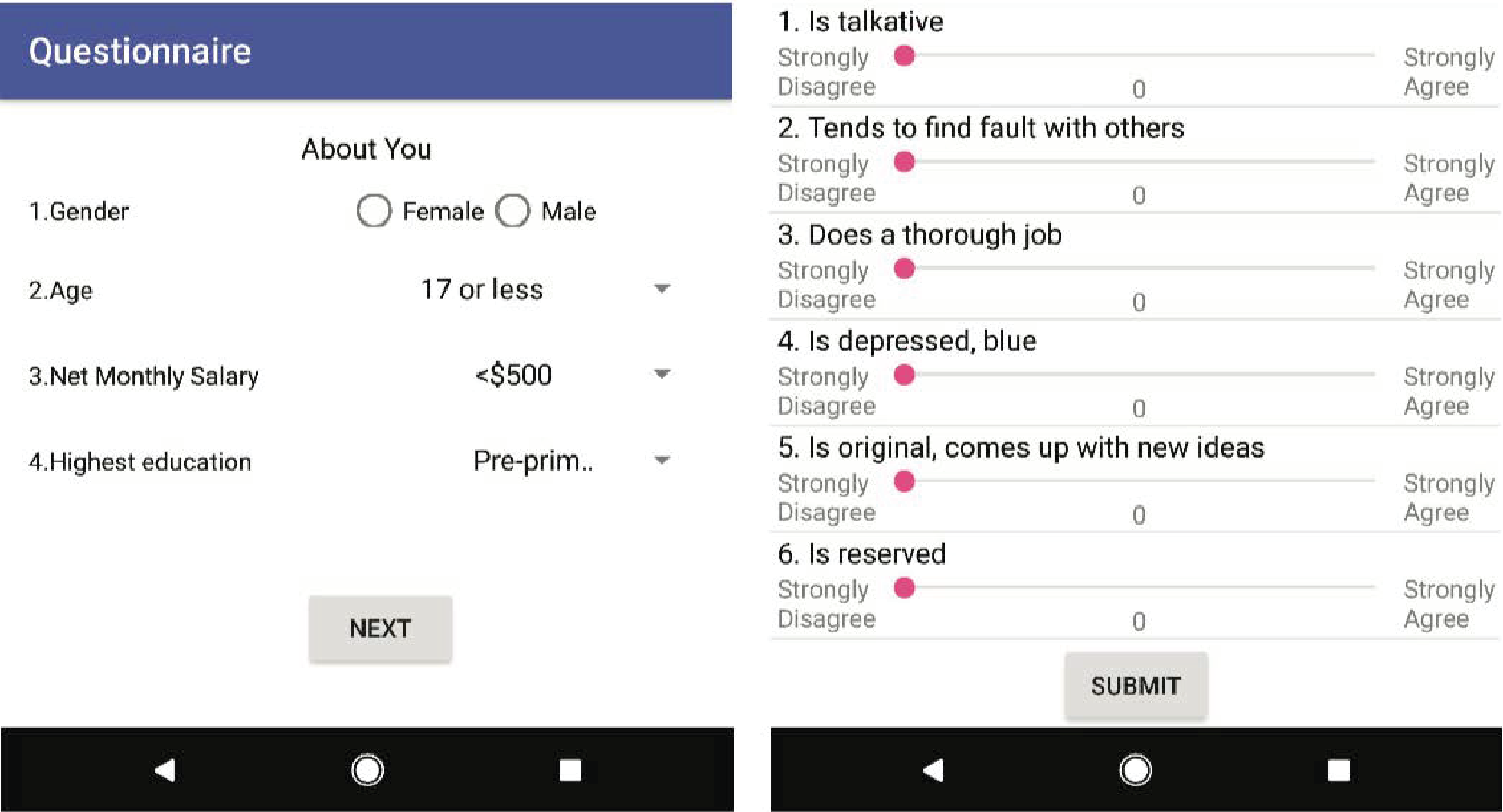}
\caption{The layouts of our designed survey app. The left part is about demographics, and the right part consists of BFI 44 questions \cite{john1999big}.}
\label{f_surveyApp}
\end{figure}

\subsection{Personality Trait Classifier} 
We generate a learning-based algorithm to predict users' personality traits from their mobile behaviours (app adoption in different categories). Inspired by prior studies \cite{seneviratne2014predicting,xu2016understanding}, we apply the information on mobile app adoption, more specifically, the numbers of installed apps on the eight different categories listed in Section \ref{ss_permissionSorting}. An extra category is added for undefined apps. It is worth noting that pre-installed (default) apps in mobile devices are excluded in our experiment as they are not indicative of a user's personality. We use these features since they are readily accessible and are proved to be useful indicators in previous data-driven studies \cite{seneviratne2014predicting,xu2016understanding}. We mainly focus on `High' and `Low' groups prediction in each personality trait, as people in those groups behave differently \cite{xu2016understanding}. In total, we generate ten models to predict the identified groups simultaneously. A series of machine learning algorithms are employed such as Random Forest, Decision Tree, SVM and Naive Bayes \cite{goldberg1988genetic}.

\subsubsection{Data Collection.} We design a questionnaire in two versions to collect real-world user information, including demographics, BFI 44 questions, and app adoption behaviours (excluding pre-installed apps). One version is an online questionnaire which requests participants to input their information manually. However, counting app numbers in several different categories is error-prone. Therefore, we develop another Android survey app to count those numbers automatically in the background. The snapshots of the app are depicted in Figure \ref{f_surveyApp}. The demographics and BFI questions are designed separately into the two layouts in the figure. The quantity of apps in each category is retrieved by Android API and sent to our server automatically along with the input of two layouts. The app is published in the Android app market for participants to download. As the functionality of returning app category is only supported in Android 5.0 (API level 21) and higher, we keep both versions of the questionnaire in our experiment.

We publish the links of our questionnaire in two separate versions on the Amazon's Mechanical Turk (MTurk) platform \cite{MTurk}. We only recruit participants with a `Masters Qualification', which is a performance measure defined by MTurk and is strictly granted to high-performance workers by MTurk. Each participant is rewarded 0.5 dollar for completing the questionnaire. We also require them to provide a survey code which is only generated after the questionnaire is completed for data validation. In total, we obtained 587 responses. We further manually removed five invalid responses submitted in a short time or with multiple duplicate answers. From the demographics statistics, we found all of the participants are adults (18 years of age or older), with gender balance (52.7\% Male and 47.3\% Female). We label the data according to the standard explained in Section \ref{ss_personality}. It is found that for each trait, around half of the participants belong to the group 'Medium', while the other two groups account for about a quarter of the participants respectively.

\subsubsection{Evaluation.} 
We evaluate the performance of our model, and further compare the result with a random model. Intuitively, the model randomly assigns a group to each user in five traits independently. In comparison, we train and test several algorithms and select the one with the best performance as our model. We achieve ten predictive models based on the process. Each model is named using a target group, such that `E-High' represents the model to predict the users perform as high in Extraversion. The performance is measured using Precision, which is the proportion of the participants who are correctly allocated in each target group. We only apply Precision to measure the prediction performance on the target group, since other measurements such as Accuracy can only provide overall performance on both groups.

\begin{figure}[t]
\centering
\includegraphics[width=1\linewidth]{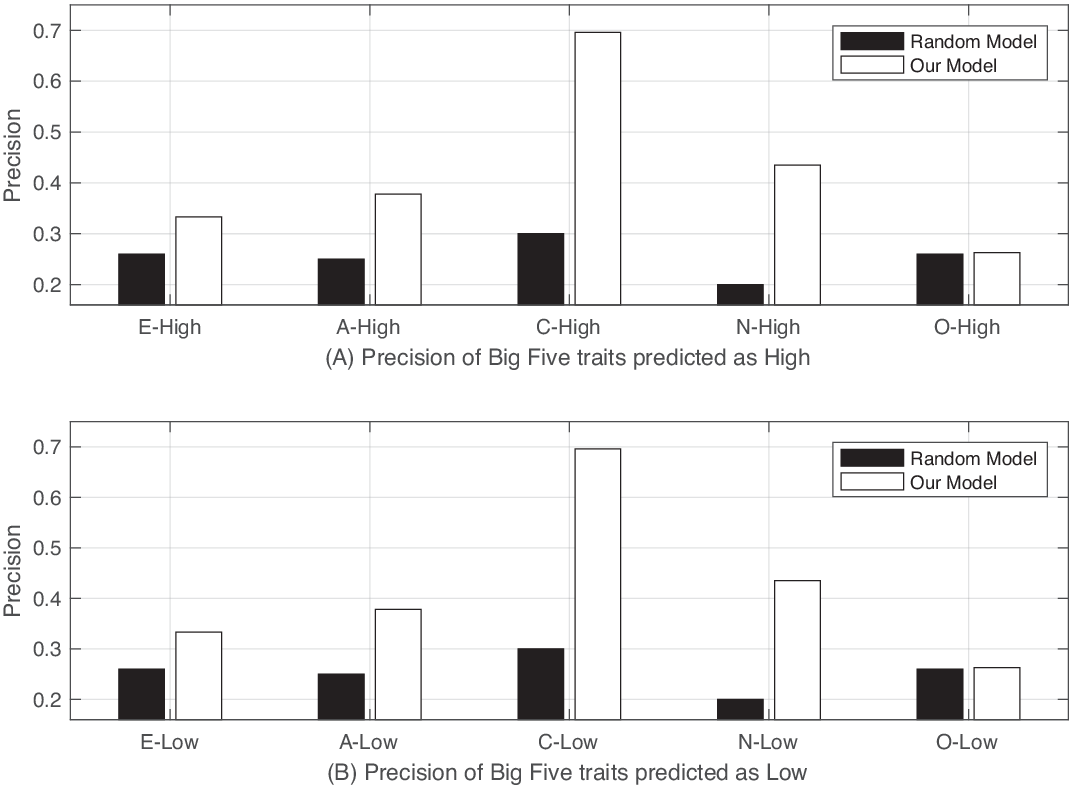}
\caption{Precision of random models and our models in Big Five traits prediction as (A) High and (B) Low.}
\label{f_personalityprecision}
\end{figure}

The comparison results between our 10 models and the random models are demonstrated in Figure \ref{f_personalityprecision}, with two sets of five-trait predictive models in group (A) `High' and (B) `Low'. The Precision of our models is higher than random models, the difference is most significant on both models of C (Conscientiousness), where it achieves approximately 40\% improvement. Other models can also increase Precision by about 15\% on average. Moreover, the performance of our model is comparable to prior similar studies \cite{wright2014personality,xu2016understanding}.

\subsection{Personality Trait and Linguistic Preference.} 
We further study the linguistic preference for predicted personality traits. An example of adjectives associated with each trait is shown in Table \ref{t_adjective} from the prior psychological study \cite{mairesse2008trainable}. We accordingly design different lexical choices with recognisable linguistic variation. For instance, extravert people are likely to use more positive emotion words compared to introvert ones. A detailed model for personalised descriptions generation is explained in section \ref{s_NLG}.

\begin{table}[t]
\caption{Example adjectives corresponding to high or low level in Big Five traits \cite{mairesse2008trainable}.}
\footnotesize
\label{t_adjective}
\centering
\begin{tabular}{c | l | l}
\hline
\textbf{Trait} & \textbf{High} & \textbf{Low} \\ \hline
E & warm, sociable, active & shy, quiet, passive \\ 
A & calm, peaceful, confident & neurotic, anxious, depressed \\ 
C & trustworthy, considerate, helpful & selfish, suspicious, malicious \\ 
N & competent, disciplined, dutiful & disorganised, impulsive, forgetful \\ 
O & creative, curious, cultured & conservative, ignorant, simple \\  
\hline
\end{tabular}
\end{table}

\subsection{Proof-of-Concept: Part II}
\begin{figure}[t]
\centering
\includegraphics[width=1\linewidth]{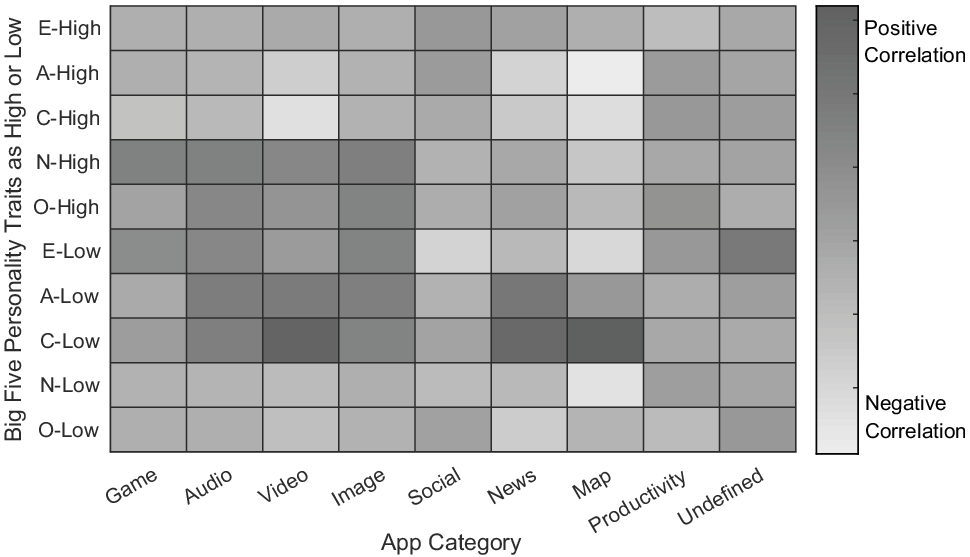}
\caption{Correlations between users' app adoption in different categories and their Big Five Personality traits in group `High' or `Low'.}
\label{f_correlation}
\end{figure}

We analyse the correlations between the features (app categories) and target groups (`High' and `Low' in five traits). Figure \ref{f_correlation} depicts Pearson correlation coefficients for pairs of features and groups, which are represented by the grayscale of the blocks. Dark blocks indicate positive correlations, while light blocks indicate negative correlations. For instance, according to the figure, users in group `C-Low' (`Low' group of Conscientiousness) are more likely to install apps in Video, News and Map. Similarly, app adoption on Game, Audio, Video and Image can denote a user in group `N-High' (`High' group of Neuroticism). In contrast, agreeable and conscientious users are in low possibility of adopting Map apps as well as the users in group `E-Low' and `N-Low' (`Low' groups of Conscientiousness and Neuroticism). The negative correlation also appears when conscientious users adopt Video apps.

\section{NLG-based Context Construction}
\label{s_NLG}
We apply a statistical model to estimate linguistic parameters to generate different syntactic templates and transform them into security-centric descriptions for different personality traits. 

\subsection{Syntactic Structure Generation}
Syntactic structures for different personality traits are generated based on a SNLG (Statistical Natural Language Generation) model. We leverage 67 parameters \cite{mairesse2007personage} in syntactic structure identification which are explained in Table \ref{t_parameters}. The parameters are estimated by several statistical models such as SVM \cite{mairesse2008trainable}. The kernel of the method is implemented using generic lexical resources including WordNet \cite{miller1995wordnet}. This provides a potential to create various personality-focused text with different kinds of content plans. Hence, we have adopted the model for security description generation based on personality traits.

As our research goal is to improve the security awareness of end users, we utilised the model to generate utterance-based app descriptions which are more natural and expressive than simple plain-sentence descriptions (\textit{e.g.}, `Err... sending SMS messages is suspicious.' v.s. `App sends SMS messages.'). We keep the kernel of the model with input of our target content and calculated personality score. The content is collected and organised from Drebin \cite{arp2014drebin} results. More specifically, we generate all malware-indicative features from their dataset including 5,560 malicious apps in the real world. Those features are then classified into several categories such as permission, according to their functionality and triggered condition. Each feature is interpreted to be a subject in one description according to corresponding pre-defined explanations in Android Document \cite{Android-doc}. Take `SEND\_SMS' as an example, it is transformed as `sending SMS messages' to insert into related scripts. The category is defined as subject type, and is likely to appear in a designed sentence (\textit{e.g.}, The  permission is suspicious). After retrieval from Android Document \cite{Android-doc}, it is discovered that the categories cover all possible features for current apps. Therefore, our method is comprehensive in description generation for both known and unknown malicious apps.

\begin{table*}[t]
\caption{The 67 parameters learned to determine specific syntactic structure, proposed from previous work \cite{mairesse2007personage}.}
\footnotesize
\label{t_parameters}
\centering
\begin{tabular}{ l | p{14cm} }
\hline
\textbf{Parameters} & \textbf{Descriptions}\\ \hline  \hline
\multicolumn{2}{l}{\textbf{Content parameters:}}\\
VERBOSITY & Control the number of propositions in the utterance \\
RESTATEMENTS & Paraphrase an existing proposition \\
REPETITIONS & Repeat an existing proposition \\
CONTENT POLARITY & Control the polarity of the propositions expressed, i.e. referring to negative or positive attributes \\
REPETITIONS POLARITY & Control the polarity of the restated propositions \\
CONCESSIONS &  Emphasise one attribute over another\\
CONCESSIONS POLARITY & Determine whether positive or negative attributes are emphasised \\ 
POLARISATION & Control whether the expressed polarity is neutral or extreme \\ 
POSITIVE CONTENT FIRST & Determine whether positive propositions-including the claim-are uttered first \\ \hline
\multicolumn{2}{l}{\textbf{Syntactic template selection parameters:}}\\
SELF-REFERENCES & Control the number of first person pronouns \\
CLAIM COMPLEXITY & Control the syntactic complexity (syntactic embedding) \\
CLAIM POLARITY & Control the connotation of the claim, i.e. whether positive or negative affect is expressed \\ \hline
\multicolumn{2}{l}{\textbf{Aggregation operations:}}\\
PERIOD & Leave two propositions in their own sentences \\
RELATIVE CLAUSE & Aggregate propositions with a relative clause \\
WITH CUE WORD & Aggregate propositions using with \\
CONJUNCTION & Join two propositions using a conjunction, or a comma if more than two propositions \\
MERGE & Merge the subject and verb of two propositions \\
ALSO CUE WORD & Join two propositions using also \\
CONTRAST - CUE WORD  & Contrast two propositions using while, but, however, on the other hand \\
JUSTIFY - CUE WORD & Justify a proposition using because, since, so \\
CONCEDE - CUE WORD & Concede a proposition using although, even if, but/though \\
MERGE WITH COMMA & Restate a proposition by repeating only the object \\
CONJ. WITH ELLIPSIS & Restate a proposition after replacing its object by an ellipsis \\ \hline
\multicolumn{2}{l}{\textbf{Pragmatic markers:}}\\
SUBJECT IMPLICITNESS  & Make the permission implicit by moving the attribute to the subject \\
NEGATION & Negate a verb by replacing its modifier by its antonym \\
SOFTENER HEDGES & Insert syntactic elements (sort of, kind of, somewhat, quite, around, rather, I think that, it seems that, it seems
to me that) to mitigate the strength of a proposition,  \textit{e.g.}, `It seems
to me that sending SMS messages doesn't have security promise.' \\
EMPHASIZER HEDGES & Insert syntactic elements (really, basically, actually, just) to strengthen a proposition, \textit{e.g.}, `Basically, sending SMS messages doesn't have security promise.' \\
ACKNOWLEDGMENTS & Insert an initial back-channel (yeah, right, ok, I see, oh, well) \\
FILLED PAUSES & Insert syntactic elements expressing hesitancy (like, I mean, err, mmhm, you know) \\
EXCLAMATION & Insert an exclamation mark \\
EXPLETIVES & Insert a swear word \\
NEAR-EXPLETIVES & Insert a near-swear word \\
COMPETENCE MITIGATION & Express the speaker's negative appraisal of the hearer's request \\
TAG QUESTION & Insert a tag question \\
STUTTERING & Duplicate the first letters of a permission's name \\
CONFIRMATION & Begin the utterance with a confirmation of the restaurant's name \\
INITIAL REJECTION  & Begin the utterance with a mild rejection, \textit{e.g.}, `I'm not sure' \\
IN-GROUP MARKER & Refer to the hearer as a member of the same social group, \textit{e.g.}, pal, mate and buddy \\
PRONOMINALIZATION & Replace occurrences of the permission's name by pronouns \\ \hline
\multicolumn{2}{l}{\textbf{Lexical choice parameters:}}\\
LEXICAL FREQUENCY & Control the average frequency of use of each content word, according to BNC frequency counts \\
WORD LENGTH & Control the average number of letters of each content word \\
VERB STRENGTH & Control the strength of the selected verbs, \textit{e.g.}, `I would suggest' vs. `I would recommend' \\
\hline
\end{tabular}
\end{table*}

\subsection{Sentence Organisation and Aggregation}
\begin{figure*}[t]
\centering
\includegraphics[width=0.9\linewidth]{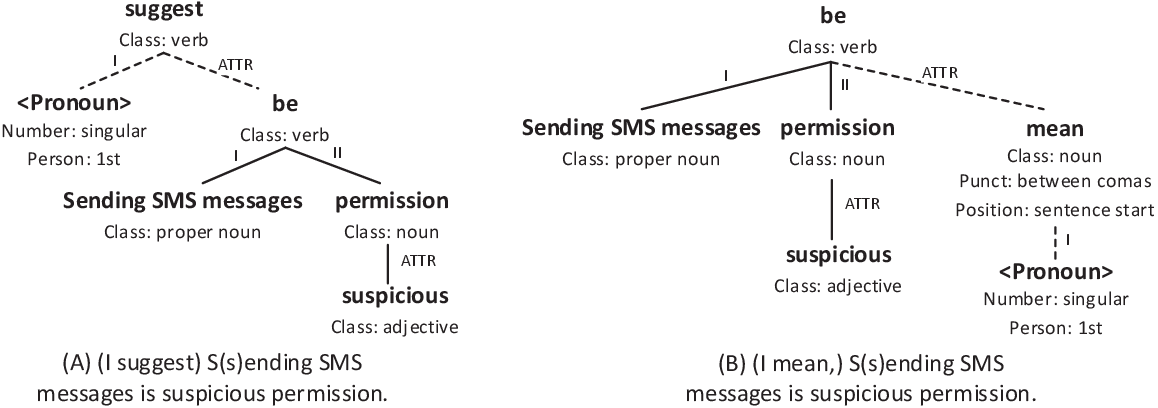}
\caption{Two example DSyntS for our security-centric descriptions. Bold text represents lexemes and variables below represent assigned attributes for RealPro at generation time. I and II indicate subject and object separately, while ATTR means modifier.}
\label{f_DSyntS}
\end{figure*}

\textbf{Sentence Organisation.} We construct the syntactic templates and convert them into text with a series of linguistic parameters. Different from previous works \cite{arp2014drebin,zhang2015towards}, we applied Deep Syntactic Structures (DSyntS) \cite{mann1988rhetorical} to generate syntactic templates. DSyntS are syntactic representations demonstrating various linguistic frameworks. Two example SyntS are shown in Figure \ref{f_DSyntS}. In the process, language for description generation is learned through multiple rule-based steps, and transformed text for semantics is modelled systematically. DSyntS produce attributes for sentence construction, such as object name (\textit{e.g.}, sending SMS messages) and object type (\textit{e.g.}, permission). RealPro surface realiser \cite{lavoie1997fast} is employed to interpret DSyntS into text representation. 

As shown in the second part of Table \ref{t_parameters}, three parameters are selected to determine syntactic templates: self-reference, complexity and polarity. Figure \ref{f_DSyntS} depicts two DSyntS examples describing the permission `sending SMS messages'. It is depicted in Figure \ref{f_DSyntS}A that the sentence is realised in two versions: basic and extended (with dashed-line branches) templates. The structure is determined by a user's personality. The introvert is found to prefer complex structure. The extended templates are provided for this group since they contain complex syntactic with a subordinate clause. For example, self-reference is related to the groups of `High' in Extraversion and Neuroticism. The extended template performs higher in self-reference parameter, with one first person pronoun compared to the lack of the pronoun in the basic template. The last parameter polarity characterises the syntactic as positive or negative. Both templates in the example exhibit the same polarity, which can be decreased by replacing `is' with `could be' in original sentences.

\textbf{Sentence Aggregation.}
After syntactic templates are generated, we further aggregate the templates to include all content and realise the combined DSyntS to one app description. Aggregation operations are determined by Aggregation parameters which are listed in the third part of Table \ref{t_parameters}. We explain how we adopt the parameters as follow. Sibling propositions from the content are linked with expression of rhetorical relation through clause-combining operation. For instance, the propositions `prevalent permission' and `at high security risk' are contrasted, so they are combined with cue words like `however' or `but'. The operation terms are selected based on possibility distribution of relation corpus, which is learned from personality parameters. The associated pairs of propositions in children nodes are aggregated by all possible clause-combining operations traversal. To achieve this, the SPARKY clause-combining operations \cite{stent2004trainable} are applied for representing rhetorical relations, as shown in Table \ref{t_clause}. Take `MERGE' as an example, it is used when the main verb for two DSyntS is same. An example result of `WITH CUE WORD' is `Sending SMS messages is the suspicious permission, with no security promise'.

Pragmatic makers do not contribute to syntactic structure, but instead, they play an important role in personalising sentences with pragmatic effects. We manually assign syntactic variables for the markers collected and combine them with the handcrafted database \cite{mairesse2008trainable}. Insertion points are detected by traversing all DSyntS, and markers are inserted correspondingly. Figure \ref{f_DSyntS}B gives an example to explain the insertion process. The root node is an insertion point which matches the syntactic pattern of the marker `I mean', so the branch is inserted under the root node (dashed lines). The attributes are specified below the lexemes, such that Position variable indicates the marker added at the start of the sentence. Adopted pragmatic markers are described in the fourth part of Table \ref{t_parameters}. The insertion process contains a random selection from syntactic patterns in pragmatic markers and matching with aggregated DSyntS. As the extravert is more likely to use implicit words, SUBJECT IMPLICITNESS parameters are assigned for the group. This can lead to the transformation from `Sending SMS messages is the suspicious permission' to `the permission is suspicious'.  

\begin{table}[t]
\caption{Clause combining operations for different rhetorical relations \cite{stent2004trainable,walker2007individual}.}
\small
\label{t_clause}
\centering
\begin{tabular}{ l | p{6cm}}
\hline
{\bf RST relation} &    {\bf Aggregation operations}\\ \hline \hline
   {\sc justify}   &       {\sc with cue word}, {\sc relative clause}, {\sc so cue word}, {\sc because cue word}, {\sc since cue word}, {\sc period} \\ \hline
   {\sc contrast}  &       {\sc merge}, {\sc however cue word}, {\sc while cue word}, {\sc conjunction}, {\sc but cue word}, {\sc on the other hand cue word}, {\sc period} \\ \hline
   {\sc infer}     &       {\sc merge}, {\sc with cue word}, {\sc relative clause}, {\sc also cue word}, {\sc conjunction}, {\sc period} \\ \hline
   {\sc concede}   &        {\sc even
   if cue word}, {\sc although cue word}, {\sc but/though cue word}\\ \hline
   {\sc restate}   &       {\sc conjunction}, {\sc merge with comma}, {\sc object ellipsis}\\ \hline

\hline
\end{tabular}
\end{table}

\subsection{Proof-of-Concept: Part III}
The different linguistic stylistic variations are presented in Table \ref{t_descriptions}. Given the content, the descriptions are generated automatically according to specific personality pattern. We adjust different input parameters to produce several sentences describing one content (sending SMS messages). Each time we select two dimensions and set values to high or low separately, with neutral values for the other traits. The linguistic diversity is noticeable in different traits of users even when conveying the same meaning. For instance, an exclamation mark is applied to project high extraversion. In contrast, the pause `Err' indicates intraversion as well as low agreeableness, which is also expressed by negative propositions. Openness to experience is related to some special words, such as `damn' (group `High') and `bloody (group `Low')' in this example.

\begin{table*}[t]
\caption{Example personalised descriptions with different personality traits. H represents group `High', and L represents group `Low' in individual dimension. The blank indicates the neutral values between H and L. }
\small
\label{t_descriptions}
\centering
\begin{tabular}{l c c c c | p{14.5cm}}
\hline
\textbf{E} & \textbf{A} & \textbf{C} & \textbf{N} & \textbf{O} & \textbf{Descriptions} \\ \hline  \hline
H & H &  &  &  & I mean, sending SMS messages doesn't have any security promise and this request is like, the suspicious permission at high risk! \\ \hline
H &  & H &  &  & Sending SMS messages is at high risk. It's the suspicious permission. This permission doesn't have any security promise. \\ \hline
 & L &  &  & H & Sending SMS messages is at high risk. Err... basically, it's the damn suspicious permission. This permission doesn't have any security promise, does it? \\ \hline
L & L &  &  &  & Sending SMS messages doesn't have any security promise, does it? Err... it is the suspicious permission. \\ \hline
L &  &  &  & L & Err... sending SMS messages is the suspicious permission. It's kind of at high risk. \\ \hline
 & L &  &  & L & Err... sending SMS messages is at high risk. It's the bloody suspicious permission. This permission doesn't have any security promise, does it? \\ 
\hline
\end{tabular}
\end{table*}

\section{Evaluation}
\label{s_evaluation}
\subsection{Semantics Correctness Checking}
We evaluate the semantics correctness by the similarity calculation between Drebin descriptions and our descriptions. To obtain the score, we employ three measures: Dandelion \cite{DandelionAPI}, ADW \cite{pilehvar2013align} and LSA \cite{landauer1998introduction}. The scale of the score is from 0 to 1, while high percentage indicates high similarity. We apply four distinct personalised descriptions to compute their similarities with Drebin description based on the same feature (sending SMS messages). The similarity scores are listed at the top part of Table \ref{t_similarity}, and all of them are higher than 70\%. Furthermore, our descriptions perform best similarities, the scores of them achieve up to 83\%. In addition, syntactic similarities are also evaluated, as shown at the bottom of Table \ref{t_similarity}. It is revealed that the main content of the sentences is not modified while using different linguistic styles with only around 55\% syntactic similarities.

\begin{table}[t]
\caption{Semantic and syntactic similarities between Drebin description and our 4 descriptions (D1-D4).}
\small
\label{t_similarity}
\centering
\begin{tabular}{l | l c c c c}
\hline
 & Measures & D1 & D2 & D3 & D4 \\ \hline
\multirow{3}{*}{Semantic} & Dandelion & 0.83 & 0.83 & 0.83 & 0.83 \\
& ADW & 0.80 & 0.72 & 0.74 & 0.76 \\
& LSA & 0.73 & 0.77 & 0.77 & 0.73 \\ \hline
Syntactic & Dandelion & 0.57 & 0.54 & 0.54 & 0.53 \\
\hline
\end{tabular}
\end{table}

\subsection{Readability Score}
\begin{figure*}[t]
\centering
\includegraphics[width=0.9\linewidth]{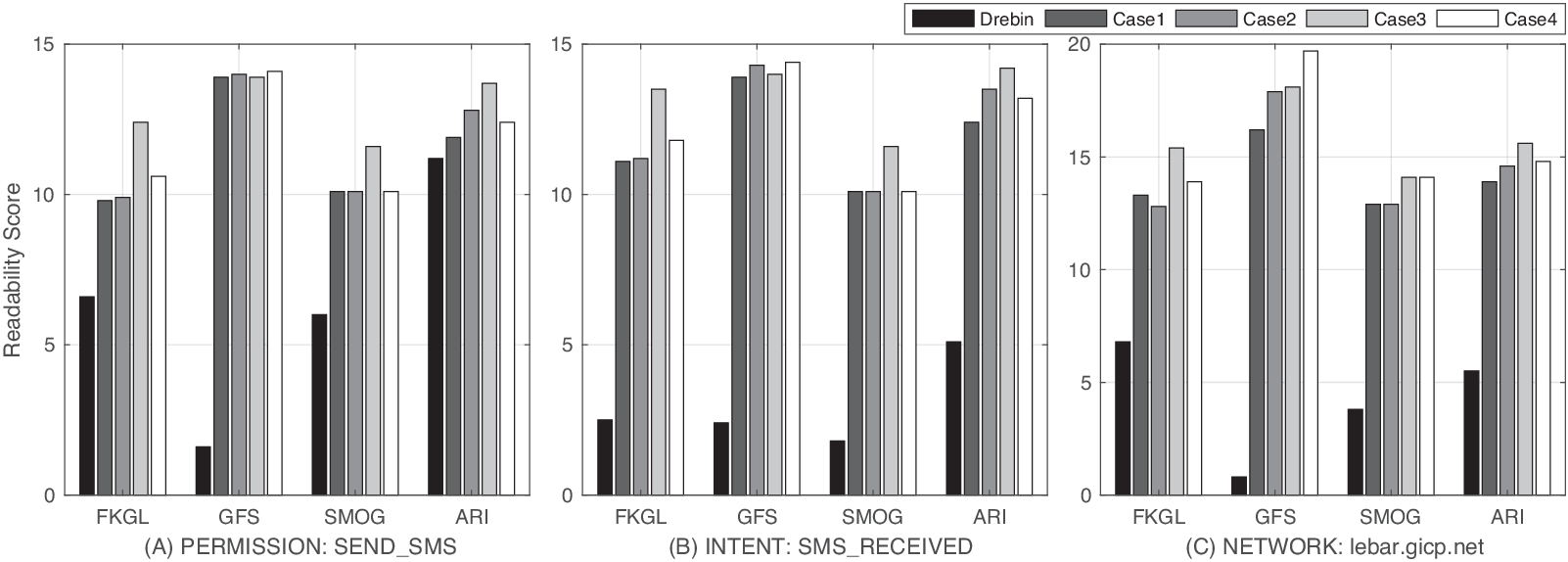}
\caption{Readability scores for Drebin description and our descriptions (Case 1-4) for three example features: (A) Permission, (B) Intent and (C) Network. We randomly select four typical personality types: Case1 (neutral values on five dimensions (E, A, C, N, O), Case2 (high values on E and C, others are neutrals), Case3 (low values on E and A, others are neutrals) and Case4 (high values on E and A, others are neutrals). Four readability indices \cite{kincaid1975derivation} are applied: FKGL (Flesch Kincaid Reading Ease), GFS (Gunning Fog Score), SMOG (SMOG Index) and ARI (Automated Readability Index).}
\label{f_RS}
\end{figure*}

To evaluate the readability of our generated descriptions, we leverage the readability score based on four indices \cite{kincaid1975derivation}: FKGL (Flesch Kincaid Reading Ease), GFS (Gunning Fog Score), SMOG (SMOG Index) and ARI (Automated Readability Index). High score means the text is easily readable, and vice versa. We apply malware-indicative features used by Drebin (Figure \ref{f_ExampleDescription}C), and for each of them, we generate four descriptions with different personality input in comparison to the original description provided by Drebin. As depicted in Figure \ref{f_RS}, each subplot represents a feature, in which 1 represents the Drebin description and Case 1-4 represent our four descriptions. It can be found that our descriptions consistently performs better than Drebin in readability regardless of which index is adopted. The score increase is most noticeable when using FKGL, which reaches to almost 20 in Figure \ref{f_RS}C.

\subsection{Readability and Awareness Checking}

We further evaluate the readability and measure the awareness improvement in real practice.
We develop an Android app including personality test (the right layout of Figure \ref{f_surveyApp}) and scenario-simulated test. Participants are provided with an Android mobile phone with our test app installed. Firstly, participants are asked to complete a personality test in the app. Once the test is finished, the user's personality type is identified. Participants are then required to download apps through our simulation platform, which is similar to the process of users are browsing and downloading apps on Google Play. After participants click the download button, a page including security descriptions is presented before installation. We provide three typical types of security descriptions on our simulation platform, including our two personalised descriptions as well as Drebin \cite{arp2014drebin} description. Our descriptions are based on the personality of each participant, where one accords with the participant's personality while the other is opposite. After participants make their decisions, our app requires participants to summarise the security descriptions and provide reasons if they choose not to download the app. 
In our experiment, the security description is regarded as readable, if participants are able to correctly summarise the security descriptions. The improvement of awareness is considered as true only if participants choose not to install the app and give the reason as `security problem'.

Before we start the experiment, we carried out a pilot study to confirm all the questions included in our experiment are all meaningful and explicitly expressed. 
Based on the pilot study, we improved our experiment from the following aspects. Participants are required to download a certain type of Android apps to avoid the impact of their preferences on some specific types.
Our app utilises similar interfaces of Google Play, positioning participants in a real app installation. 
We also avoid other security or any risk related hints during the experiment, to ensure that the security awareness only comes from the descriptions we generated. We adjust the logical order of the page processing in our test app to achieve user-friendly interaction.


In total, we invited 38 participants, and among all the participants, only 2 participants guessed our research objectives during the experiment. It indicates that apart from these 2 participants, rather than being potentially influenced by our research objectives, the rest of participants perform naturally by following their own instincts and their daily behaviours of downloading apps. Besides, we found that education background is a critical impact factor in our experiment. We invited 12 students who major in IT security, the result indicates that only one of them chose to install the app with the security descriptions. The reason was that he regarded the app as a low risk and it was acceptable to obtain its benefit. Since our descriptions are designed for ordinary users, and the majority of them do not have related knowledge, we excluded those IT students from our experiment. Therefore, only remaining 24 responses are regarded as valid.

From our experiment, the readability of our personalised descriptions which accord with participants' personalities achieved the best result with 79.2\% participants correctly summarised the main idea of the descriptions, followed by the descriptions opposite to participants' personalities, and the percentage is 66.7\%. In comparison, the readability of the description from Drebin only reached 41.7\%. In terms of security awareness improvement, our descriptions also achieved better result than Drebin description.
The descriptions according with participants' personalities 
achieved 54.2\%. 
In contrast, Drebin only achieved 29.2\%. 



\subsection{System Efficiency Checking}
We evaluate our system efficiency based on runtime performance of randomly selected 150 apps from the Google Play Store. Our system contains four parts: Drebin detection model \cite{arp2014drebin}, permission ranking, personality classifier and NLG model. While the testing experiments run fast, the training process is time consuming. Fortunately, model training is one-time effort, and is generally finished offline before use. Three models involved are trained on a PC (2.7 GHz Inter Core i5 with 8GB 1867 MHz DDR3). Drebin \cite{arp2014drebin} model is generated in 63 minutes including the app analysis time, with their whole dataset consisting of 131,611 apps. In contrast, the processes in personality classifier and NLG model are more efficient, completed in 5 seconds and 18 minutes respectively with 586 responses and 8 sets containing 545,334 potentially suspicious features from detection model training result.

We test the four parts individually on the PC, and we further run the complete app on two Android smartphones (Nexus 5 and Pixel XL). The size of the apps we collect is ranged from around 50 KB to 5MB. The average testing time on PC for an app is around 1,475ms (700 ms, 370 ms, 150 ms and 255 ms for four parts respectively). The whole process costs less than 2 seconds per app including file analysis (about 280 ms). The result is comparable to DescribeMe \cite{zhang2015towards}, with average runtime in 391.5 seconds. We further tested the prototype in two smartphones, and both of them achieved the results in 10 seconds.

\section{Discussion}
\label{s_dis}

\subsection{Dependence of Detection Accuracy.}
PERSCRIPTION is implemented based on the malware detection method from Drebin \cite{arp2014drebin}. Since all of our generated descriptions originate from trained malware indicative features, the detection performance considerably contributes to the explanation correctness. Although the classifier was evaluated to outperform existing methods and anti-virus scanners, there is still room to improve the detection rate. As suspicious apps are evolving rapidly, detection model is required to update frequently.

\subsection{Dependence of Personality Judgement.}
The linguistic preference and syntactic structure are generated according to judged personality traits. Although the Big Five model is a widely accepted measurement to distinguish human personality, it is still challenging to identify a user trait exactly in five dimensions. Moreover, there may exist improvement in our personality predictive model based on user app adoption. Some features may be more personality indicative and thus have the potential to improve the model accuracy such as sensitive information such as SMS message length or content. However, we do not consider those features in our automatic system because they are not readily accessible. In practice, from the investigation of our user study, users tend to protect their private information from external access.

\subsection{Dependence of NLG Performance.}
Unlike the traditional sentence template, our personalised descriptions are generated automatically with NLG technique. Currently, our work employs the model for text generation trained by previous paper \cite{mairesse2007personage} which demonstrates stylistic variations in different personality traits. While the model produces distinct generic linguistic preference to improve text readability, more generation styles may be considered to target users' security concerns. Some technical words in descriptions may also be interpreted for comprehension towards average users. Moreover, when evaluating users' security awareness improvement, we also provided participants with the descriptions opposite to their personalities. The result (50\%) is not significantly worse compared to the descriptions according with their personalities (54.2\%). This can result from the small sample size of the participants. More importantly, we provided participants with several descriptions for the same content, once the first description is presented, they may get preconceptions on the content. This can influence their decisions with the remaining descriptions. We may leverage some experiments from psychology area to avoid the impact for future work.


\section{Related Work}
\label{s_relatedwork}

\textbf{Android App Description Generation.} A series of efforts have been made to generate descriptions for Android apps. WHYPER \cite{pandita2013whyper} was the first to utilise Natural Language Processing (NLP) techniques to explain permission requests. AutoCog \cite{qu2014autocog} further related textual descriptions with permission list via NLP and machine learning algorithms, and exposed low description-to-permission fidelity. Therefore, later studies generate descriptions from the malicious behaviours detected by code analysis. DescribeMe \cite{zhang2015towards} analysed internal program logics and translated security-sensitive code patterns into natural language scripts. Similarly, AutoPPG \cite{yu2017toward} performs static analysis and achieved app behaviours about user personal information to generate privacy policy. Drebin \cite{arp2014drebin} employed learning-based algorithms to identity indicative patterns for malware, which were embedded into constructed sentence templates. Compared with them, PERSCRIPTION mainly focuses on security awareness improvement for end users, with personalised descriptions to aid risk assessment psychologically.

\textbf{Users' Concerns Learning.} Prior works studied users' concerns from different angles. Adrienne et al. indicated that users' attention can be reflected from user views which provided privacy-related warning \cite{felt2012Android}. Some studies \cite{fisher2012short,mugan2011understandable,wilson2013privacy} focused on learning users' adoption of location-based services and their concerns which was the permission users concerned most from our study result. Later, Jialu et al. applied hierarchical clustering with an agglomerative approach to cluster user's privacy preferences on mobile app \cite{lin2014modeling}. Bin et al. characterised the concerns with a hierarchy of three privacy levels \cite{liu2015personalized}. Bin et al. proposed a method to learn privacy profiles from permission settings to assist users with further configuration \cite{liu2016follow}. As a comparison, we leveraged users' privacy concerns from the amount of apps they changed default settings instead of the status `Allow'. We found the modification behaviour was more indicative for users' concerns and used to re-organise the order of our security-centric app descriptions.

\textbf{User Trait and Linguistic Preference.} Correlations are revealed from prior studies among mobile behaviours, user traits, and linguistic preference. Suranga et al. \cite{seneviratne2014predicting} predicted user traits according to the list of app installed and achieved above 90\% precision. Another similar predictive model with high precision was proposed to classify users into Big Five personality traits \cite{john1999big} with the indicators of mobile app adoption \cite{xu2016understanding}. Moreover, Big Five personality traits are widely accepted in psychology and demonstrate linguistic variation \cite{saucier2015makes,majumder2017deep}. Fran{\c{c}}ois et al. conducted a series of studies to present the stylistic linguistic variation in different personality traits and further produce a language generator with continuous stylistic dimensions \cite{mairesse2008trainable,mairesse2010towards,mairesse2011controlling}. Inspired by these works, we combine two models to automatically provide personalised descriptions with users' linguistic preference according to the information from app installation.

\section{Conclusion}
\label{s_conclusion}
In conclusion, we propose an innovative method to automatically generate security-centric psychologically-informed description by learning users' concerns and linguistic preference. The implemented system, PERSCRIPTION, is evaluated to outperform existing description generators. Experiment results and a further user study demonstrate that our generated text is more readable and more comprehensive to cater users' attentions and concerns. And also, the system can output semantics-correct descriptive text in a couple of seconds. Therefore, PERSCRITION is effective in security awareness improvement and aiding risk assessment for end users.






\bibliographystyle{ACM-Reference-Format}
\bibliography{ccs-sample}

\end{document}